# Elucidating the photoresponse of ultrathin MoS$_2$ field-effect transistors by scanning photocurrent microscopy

Chung-Chiang Wu[1,||], Deep Jariwala[1,||], Vinod K. Sangwan[1], Tobin J. Marks[1,2], Mark C. Hersam[1,2,3], and Lincoln J. Lauhon[1,*]

[1]Department of Materials Science and Engineering, Northwestern University, Evanston, IL 60208, USA

[2]Department of Chemistry, Northwestern University, Evanston, IL 60208, USA

[3]Department of Medicine, Northwestern University, Evanston, IL 60208, USA

|| These authors contributed equally.

**Abstract:**

The mechanisms underlying the intrinsic photoresponse of few-layer (FL) molybdenum disulphide (MoS$_2$) field-effect transistors are investigated *via* scanning photocurrent microscopy. We attribute the locally enhanced photocurrent to band-bending assisted separation of photoexcited carriers at the MoS$_2$/Au interface. The wavelength-dependent photocurrents of few layer MoS$_2$ transistors qualitatively follow the optical absorption spectra of MoS$_2$, providing direct evidence of interband photoexcitation. Time and spectrally resolved photocurrent measurements at varying external electric fields and carrier concentrations establish that drift-diffusion currents dominate photothermoelectric currents in devices under bias.

TEXT:

The layered transition metal dichalcogenides (TMDCs)[1] have attracted great interest recently due to their intriguing electrical and optical properties.[2-8] Field-effect transistors (FETs) fabricated with single layer (SL) and few layer (FL) $MoS_2$ have shown both unipolar[9-11] and ambipolar[12] charge transport characteristics with high in-plane electron mobility concurrent with high on/off current ratios[9,13] and large current-carrying capacity[10]. While bulk $MoS_2$ is an n-type semiconductor with an indirect bandgap of ~1.3 eV[14], single-layer $MoS_2$ has a direct bandgap of ~1.8 eV[15,16], which leads to enhanced photoluminescence (PL) compared to bilayer and thicker samples. The combination of a large bandgap in the visible region, strong photoresponse, light emission, and high field-effect mobility makes $MoS_2$ a promising 2D semiconductor for a variety of electronic[6,17,18] and optoelectronic[7,19,20] applications. Besides the above mentioned, several other interesting properties such as piezoelectricity,[21] tunability of band gaps and phase transitions with electric field, strain and composition[22-25] have also been predicted for ultrathin TMDCs. Solution processed two dimensional materials, decorated with metal nanoparticles or quantum dots (QDs) have also been heavily investigated as catalysts and electrode materials in photochemical reactions.[26-28] The high surface area, coupled with a band gap in visible part of the electromagnetic spectrum, makes these ultrathin dichalcogenide based catalysts attractive candidates for solar water splitting.[29-32] The interface of these nanoparticles with $MoS_2$ plays a deterministic role in the charge transfer kinetics and efficiency of such photoelectrochemical reactions.[33,34] Thus, the behavior of photoexcited carriers in a controlled FET geometry is relevant to both optoelectronic and energy conversion applications.

The photoresponse of $MoS_2$-based optoelectronic devices has been attributed to various mechanisms including photoconductivity,[19] photovoltaic effects,[35] and the photothermoelectric

(PTE) effect.[36] More specifically, global illumination of MoS$_2$ FETs has produced photocurrents and open circuit voltages ascribed to photoconductivity changes[19] and photovoltaic effects,[35] respectively. In contrast, a recent study employing local illumination with a focused beam concluded that the PTE effect dominates the photoresponse of SL MoS$_2$ FETs with ohmic contacts.[36] Prior work on carbon nanotube and graphene devices has revealed a similar diversity of behaviors that can inform investigations of MoS$_2$. The PTE effect predominates in carbon nanotube FETs under some circumstances,[37-39] and the contributions of photovoltaic and thermoelectric signals in graphene have been convincingly isolated.[40-48] For example, the graphene photovoltaic response dominates the PTE response under applied bias and at continuous wave excitation at low temperatures.[40,47]

Here we employ a combination of spectroscopic and time-resolved scanning photocurrent microscopy (SPCM) measurements to identify the nature of the photoresponse of ultrathin MoS$_2$ FETs under a comprehensive range of source-drain and gate biasing conditions including depletion, accumulation, and saturation. Although we fabricated and measured devices consisting of SL, 3L and 4L MoS$_2$ flakes, we focus our analysis on 4L MoS$_2$ FETs because they can be biased into saturation in air without being damaged. We conclude that the photocurrent is dominated by drift and diffusion of photoexcited carriers in regions of high electric field, both in the device channel and adjacent to the MoS$_2$/Au contacts. The photocurrent absorption spectra in MoS$_2$ flakes of different thicknesses, the photoresponse time, and the bias dependencies of the photocurrent provide a comprehensive picture of the nature of the response.

FETs were fabricated by e-beam lithography on mechanically exfoliated MoS$_2$ flakes on n$^+$-Si/SiO$_2$ (300 nm) substrates.[11] MoS$_2$ flakes were first screened by optical contrast imaging, and the thickness of each flake was characterized by Raman spectroscopy[49] (see Supporting

Information S1). Raman spectra (Figure S1) obtained from three samples are assigned to single- (SL), triple- (3L) and quadruple- (4L) layer MoS$_2$, respectively. The peak at ~385 cm$^{-1}$ corresponds to the in plane ($E_{2g}^1$) mode, and the peak at ~404 cm$^{-1}$ is attributed to the out of plane ($A_{1g}$) mode.[49] The $E_{2g}^1$ mode softens and $A_{1g}$ mode stiffens with increasing number of layer. In this manner, the frequency difference between these two modes was employed to determine MoS$_2$ thickness.[49]

Figure 1a shows a schematic of the experimental setup and the geometry of a typical MoS$_2$ FET. A scanning confocal microscope (WiTec) coupled to a tunable coherent white light source (NKT Photonics) is used to generate the spatially- and spectrally-resolved photocurrent, which is converted into a voltage by a current preamplifier and recorded by either a lock-in amplifier (for imaging) or a digital sampling oscilloscope (for temporally resolved measurements). The reflected light is recorded simultaneously to correlate the spatial photocurrent map with the device geometry. A representative photocurrent image of a 4L device is shown in Figure 1b for zero drain-source bias, an excitation wavelength λ=550 nm, and an excitation power of 20 µW. Positive (bright) and negative (dark) photocurrent is observed near the edge of the source (S) and drain (D) contacts, respectively. Qualitatively similar behavior is also observed in the SL device (see Supporting Information S2). The transfer characteristic of the 4L MoS$_2$ FET (Figure 1c) exhibits n-type semiconducting behavior with a field-effect mobility of ~8 cm$^2$/Vs, a typical value for FL MoS$_2$ FETs on SiO$_2$ dielectrics when measured under ambient conditions without encapsulation.[50,51] The output plot of the 4L device is linear at a low drain bias (Inset in Figure 1c) in agreement with prior reports that Au forms an ohmic contact.[9,52] In addition, the output plot shows a saturation regime consistent with the formation of a pinch-off region as $V_{DS}$ exceeds $V_G-V_T$; direct evidence of pinch-off is provided below. The 4L output plot is well described by the quadratic current versus voltage relationship expected for a conventional planar FET[6], giving $V_T$ = -9.3 V. We note that linear current versus voltage characteristics can be observed even in the presence of a Schottky barrier if charge injection is dominated by tunneling.[53] This may be the case in the devices under consideration, although we also note that the channel resistance likely dominates the contact resistance.[54] As discussed further below, some degree of band bending near the contacts can be expected considering the recent report of electroluminescence at the MoS$_2$-metal interface.[55] An analysis of carrier lifetime ($\tau$) and minority carrier diffusion length ($L_D$) enabled by the SPCM measurement is presented in Supporting Information section S3.

An important goal of the SPCM characterization is to identify the dominant photocurrent generation mechanisms in different transistor biasing regimes, as such understanding can aid in the interpretation and optimization of transistor performance[56] and inform the design of photodetectors. Towards that end, we establish that the dominant photocurrent under bias results from interband absorption within the MoS$_2$, rather than absorption in the metal contacts. The constant power (20 µW) photocurrent spectra in Figures 2a are consistent with interband absorption in MoS$_2$. Ultra-thin MoS$_2$ has two absorption peaks at ~600 nm and ~660 nm due to the spin-orbital splitting of the valence band (although the lower energy peak of a 4L device is not clearly resolved). We note that wavelength-dependent thin-film interference within the underlying dielectric also influences the absorption in the MoS$_2$, but to a smaller extent. Qualitatively, the semi-log plots of photocurrent vs. excitation wavelength of SL, 3L, and 4L MoS$_2$ FETs show the thickness dependent transition from indirect to direct interband photoexcitation (Figure 2b), but the onset of the indirect gap is not sufficiently resolved to justify fitting. In further analyses and discussion below, we can safely assume that observed photocurrents originate from inter-band excitation in the $MoS_2$ rather than carrier generation in or heating of the metal contacts.

Studies of the local photocurrent in the 4L device under varying applied biases (Figure 3) were used to analyze FET performance and aide in the identification of the photocurrent generation mechanisms. As shown in the middle panel of Figure 3a, the Au contact induces upward band bending in the n-type semiconductor with an estimated Schottky barrier of ~400 meV based on the work function differences of Au and $MoS_2$ ($\Phi_{SB} = \Phi_{Au} - \Phi_{MoS_2}$).[55] Band bending induced photocurrent near contacts has been widely reported in devices based on silicon nanowires,[57,58] carbon nanotubes,[59,60] and graphene.[40,42,45,61] When $MoS_2$ is illuminated near the drain contact, the built-in electric field drives holes to the drain electrode while electrons move towards the source electrode, resulting in a negative photocurrent. The same process induces a positive photocurrent upon illumination near the source electrode. In addition, photothermoelectric (PTE) currents of the same sign are produced due to the heating of the junction produced by laser illumination.[36]

To explore the regime in which drift currents dominate PTE effects, photocurrent images were generated at moderate source-drain biases (Figure 3a). In the linear regime (0.5 V to –0.5 V, see Figure 1c inset), variations in $V_{DS}$ lead to enhancement and suppression of the photocurrent at opposite contacts, reflecting modulation of the near contact band bending (Figure 3a). A positive $V_{DS}$ increases the band bending at the source contact, resulting in an increase in the magnitude of the photocurrent, whereas the reduced band bending at the drain contact suppresses the photocurrent. The same behavior is observed when the bias is reversed, with enhancement at the drain and suppression at the source. The line profiles of photocurrent along the channel (white dash line) at $V_{DS}$= ±0.5 and 0 V indicate the magnitudes of the induced and zero bias photocurrents (Figure 3b). Similar observations on the SL FET are presented in Supporting Information S2. The photocurrent sign indicates that the applied electric field generates photocurrents that dominate contributions from the PTE effect and band bending at zero bias. The gate bias dependence of the photocurrents follows trends consistent with this interpretation. At large positive gate voltages, the Fermi level is shifted close to the conduction band, which decreases the built-in electric field near the $MoS_2$/Au interface (see Supporting Information S4). We observe a corresponding increase in the photocurrent magnitude. Depletion at negative gate voltages reduces band bending near the contacts, reducing the magnitude of the photocurrent. The decreasing channel resistance at increasing gate bias also contributes to an increase in photocurrent, as electrons must flow out the opposite contact to maintain charge neutrality.

In saturation (Figure 3c), positive photocurrent is observed throughout the device due to carrier drift generated by the large electric field. In further support of this interpretation, we identify the peak in the photocurrent at high drain-source biases (Figure 3d) with the space-charge region induced by pinch-off in the channel.[6,53] By integrating the photocurrent along the channel to generate an approximation of the local potential, one can observe that most of the applied voltage drops across the space-charge region, (dashed black line, $V_{DS}$=10 V).[56] We therefore conclude that the increase in photocurrent in the saturation regime mainly arises from more effective separation and collection of photoexcited carriers at high electric field. A recent study investigating the photocurrent in the saturation regime concludes that the generation current under global illumination is caused by the vanishing of pinch-off (in the space-charge region).[7] In the present study, the field distribution was probed directly, rather than inferred from the global photoresponse.

The dependencies of the photocurrent on gate bias and excitation power enable comparison of gate-induced and photogenerated carrier concentrations and transport mechanisms. The magnitude of the local photocurrent increases monotonically with gate voltage in a manner similar to the drain current (Figure 4a). The transition of the FET channel from accumulation to depletion is accompanied by a change in the power dependence of the photocurrent, reflecting the influence of the majority carrier concentration on photogenerated carrier recombination rates (Figure 4b). Specifically, the photocurrent becomes increasingly sub-linear as the transistor moves into depletion, reflecting the increasing rate of bimolecular recombination as the photogenerated carrier concentration exceeds the intrinsic carrier concentration. Simple estimates suggest (see Supporting Information S5) that the photogenerated carrier concentration is approximately $3\times10^{11}$ cm$^{-2}$, whereas the intrinsic carrier concentration is ~$10^{10}$ cm$^{-2}$ at $V_g$=0 V.

A recent study of ohmically contacted SL-MoS$_2$ devices attributed photocurrent generated near the contacts to the PTE effect,[36] which arises from the difference in the Seebeck coefficients of the contact and channel materials in the presence of light-induced temperature gradients. At zero bias, both PTE and band-bending induced photocurrents are expected to produce photocurrents of opposite polarity at opposite contacts, as noted above. While the PTE must make some contribution to the photocurrent, the drain-source bias dependence of our data *cannot* be explained by a PTE effect. First, it is clear that the photocurrent signal in Figures 3c,d is controlled by the applied electric field. The minority carrier diffusion length (see Supporting information S3) is too small to enable direct collection of hot carriers at the contacts, so one can interpret the local photocurrent signal as probing the local electric field in the channel.[54] Second, the bias dependence of the photocurrent near the source and drain contacts in Figure 3a varies in the manner one would expect if the currents were dominated by electric field induced carrier separation. There is likely to be a small contribution from PTE currents near the contacts, as has been observed for nanotubes[39] and graphene,[44,61] but these appear to be dominated by field effects in our devices under bias. In addition, the fast temporal response (~30 μs) observed in our devices (see Supporting information S6) rules out the PTE contribution resulting from heating of the contact junction, for which longer equilibration times are expected. Finally, the sub-linear dependence of the photocurrent (Figure 4b) rules out a PTE effect from electrons that have equilibrated with the lattice, as one would expect a linear power dependence in the signal.

In conclusion, spectrally and temporally resolved SPCM studies of $MoS_2$ FETs of varying channel thicknesses were carried out to identify the dominant photoresponse mechanisms. Photocurrents under bias are dominated by the field-assisted separation of photoexcited carriers as has been observed in graphene devices,[40,42,43] and the spectral dependence reveals the thickness-dependent transition from indirect to direct interband photoexcitation in $MoS_2$. Direct imaging of field-induced currents enables correlation of device saturation with channel pinch-off. The mechanistic understanding provided here will help guide the development of optoelectronic devices based on $MoS_2$. In addition, this further points to approaches for tailoring the photoresponse via chemical functionalization. Specifically, just as the gate voltage can modulate a photocurrent generated by drift/diffusion, the photoresponse can also be modulated by charge transfer to or from adsorbates/functional groups[62-64] or hybrid structures such as quantum dots similar to the case of graphene.[26,27,65,66]

**AUTHOR INFORMATION**

**Corresponding Author**

*Email: lauhon@northwestern.edu

**Notes**

The authors declare no competing financial interest.

**ACKNOWLEDGEMENTS**

This research was supported by the Materials Research Science and Engineering Center (MRSEC) of Northwestern University (NSF DMR-1121262). The authors thank B. Myers of NUANCE for assistance with electron beam lithography. This research made use of the NUANCE Center at Northwestern University, which is supported by NSF-NSEC, NSF-MRSEC, Keck Foundation, and the State of Illinois.

**ASSOCIATED CONTENT**

**Supporting information**

Experimental details, Raman spectroscopy, calculation of carrier lifetime and diffusion length, estimation of photoexcited and intrinsic carrier concentrations and time-resolved photoresponse. This material is free of charge via the Internet at http://pubs.acs.org.

**Figures:**

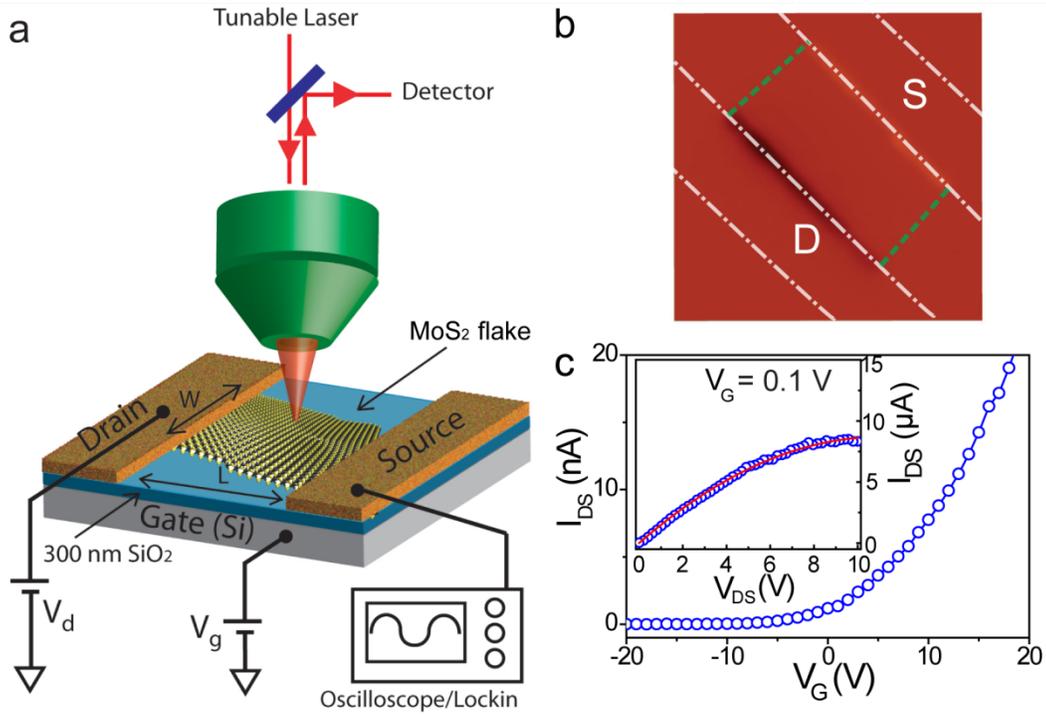

Figure 1. a) Schematic of the experimental setup for scanning photocurrent imaging of MoS$_2$ field-effect transistors (FETs). b) Scanning photocurrent image of a 4L MoS$_2$ FET acquired at $V_{DS}$, $V_G = 0$ V, and $\lambda = 550$ nm. The white dashed lines indicate the position of source (S) and drain (D) contacts (1 µm width), and the green dashed lines indicate the perimeter of the 4L MoS$_2$ flake. c) Transfer curve of a 4L MoS$_2$ FET (L = 3.5 µm, W = 6.8 µm) at $V_{DS} = 10$ mV. Inset: output curve at $V_G = 0.1$ V.

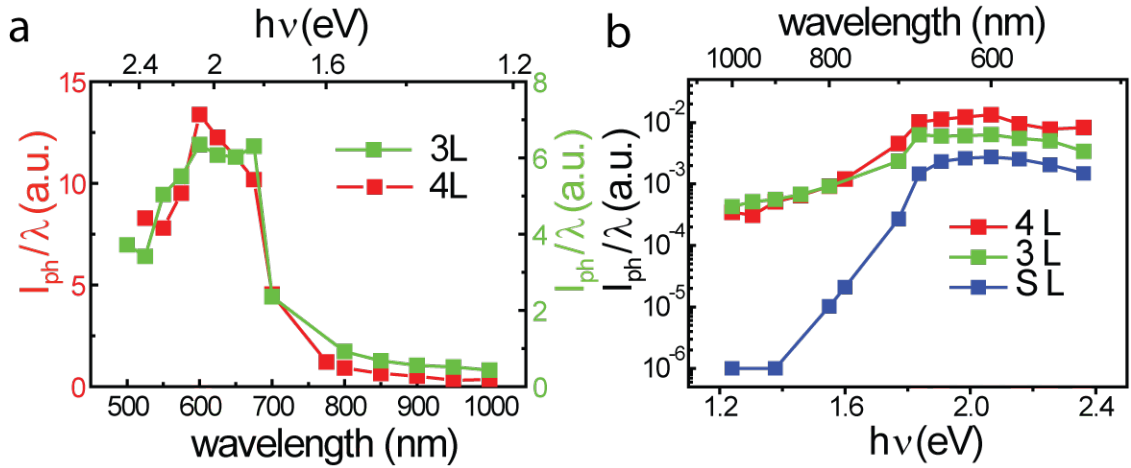

Figure 2. Photocurrent spectra at constant incident power (20 µW). To compensate for the difference in the incident photon flux, the measured photocurrents are normalized by the corresponding excitation wavelengths. a) Spectra of 3L and 4L $MoS_2$ on a linear scale. b) Comparison of SL, 3L, and 4L spectra on a semi-log scale.

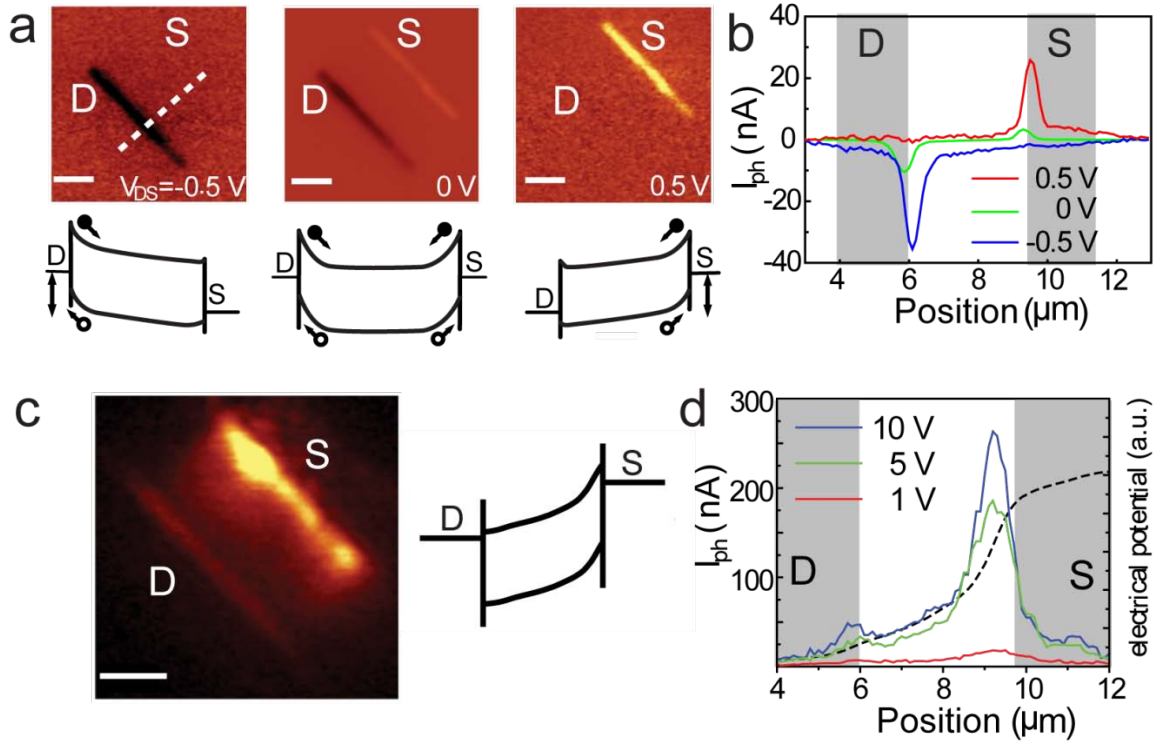

Figure 3. Drain-source bias dependence of photocurrents. a) Spatial photocurrent images in the linear region of a 4L FET at $V_{DS}$= 0.5, 0, and – 0.5 V, $V_G$=0 V shown along with the corresponding band diagrams ($V_G$ = 0 V). b) Corresponding line profiles of photocurrents along the channel (white dash line) at $V_{DS}$= 0.5, 0, and – 0.5 V c) (Left) Spatial photocurrent image recorded at $V_{DS}$=10 V and $V_G$=0.1 V (saturation regime). The channel length is 3.5 µm. (Right) Band diagram of $MoS_2$ FET in the saturation regime. d) Line profiles of photocurrents at different $V_{DS}$ and the approximate electrostatic potential along the channel (dashed black line, $V_{DS}$=10 V).

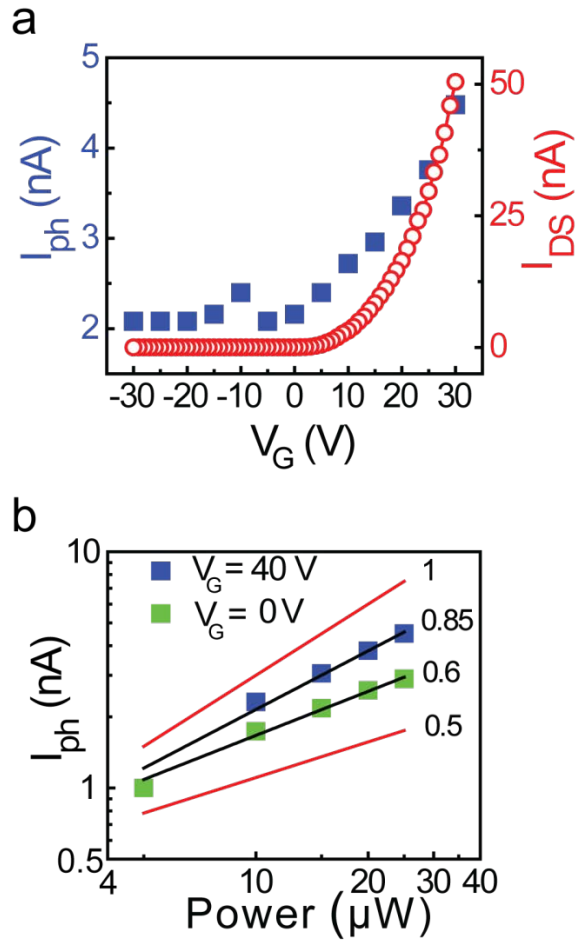

Figure 4. a) Photocurrent (blue) and drain current (red, $V_{DS}$=10 mV) as a function of $V_G$ for a 4L MoS$_2$ FET illuminated at $\lambda$=600 nm. b) Log-log plot of photocurrent amplitude as a function of excitation power ($\lambda$=600 nm) at $V_g$ = 40 V and 0 V. Linear and square root dependencies are shown as red lines for reference.

TABLE OF CONTENTS FIGURE

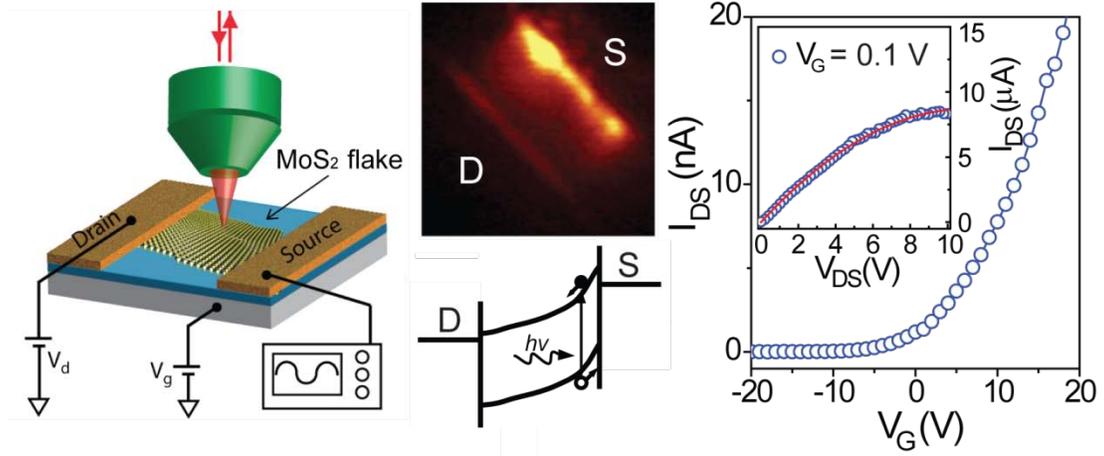

KEYWORDS: MoS$_2$, transition metal dichalcogenides, scanning photocurrent microscopy, nanoelectronics.

# Supporting Information: Elucidating the photoresponse of ultrathin MoS$_2$ field-effect transistors by scanning photocurrent microscopy

Chung-Chiang Wu[1∥], Deep Jariwala[1∥], Vinod K. Sangwan[1], Tobin J. Marks[1,2], Mark C. Hersam[1,2,3], and Lincoln J. Lauhon[1*]

[1] Department of Materials Science and Engineering, [2] Department of Chemistry, [3] Department of Medicine, Northwestern University, Evanston, IL 60208, USA
∥ These authors contributed equally.

Corresponding author:
*Email:  lauhon@northwestern.edu

## Section S1: Raman microscopy characterization of MoS$_2$ thin flakes

Raman spectroscopy was used to determine the number of layers in each of the flakes used to make devices. The Raman spectra from each sample were obtained by using a 532 nm excitation laser in a scanning confocal microscope (WITec Alpha300 R), shown in Figure S1. The peak at ~385 cm$^{-1}$ corresponds to the in plane ($E_{2g}^1$) mode, and that at ~404 cm$^{-1}$ is attributed to the out of plane ($A_{1g}$) mode. With the increase in number of layers, the $E_{2g}^1$ mode softens (the frequency decreases) and the $A_{1g}$ mode stiffens (frequency increases), leading to a monotonic increase in frequency difference between these two modes.[1] Therefore, we assigned these three MoS$_2$ flakes to single- (1L), triple- (3L) and quadruple- (4L) layer MoS$_2$ based on the frequency difference of 18.5, 23, and 24 (cm$^{-1}$), respectively. These values agree with values reported in the literature.

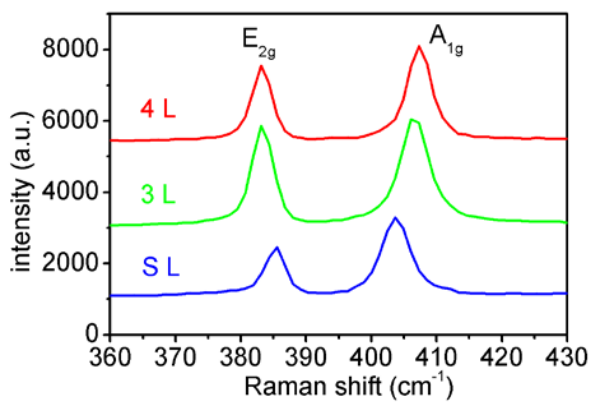

Figure S1: Raman spectra of three MoS$_2$ flakes. They are assigned to single- (SL), triple- (3L) and quadruple- (4L) layer MoS$_2$ based on the frequency difference, respectively.

**Section S2: SPCM images and electrical characteristics of single layer MoS$_2$ FET.**

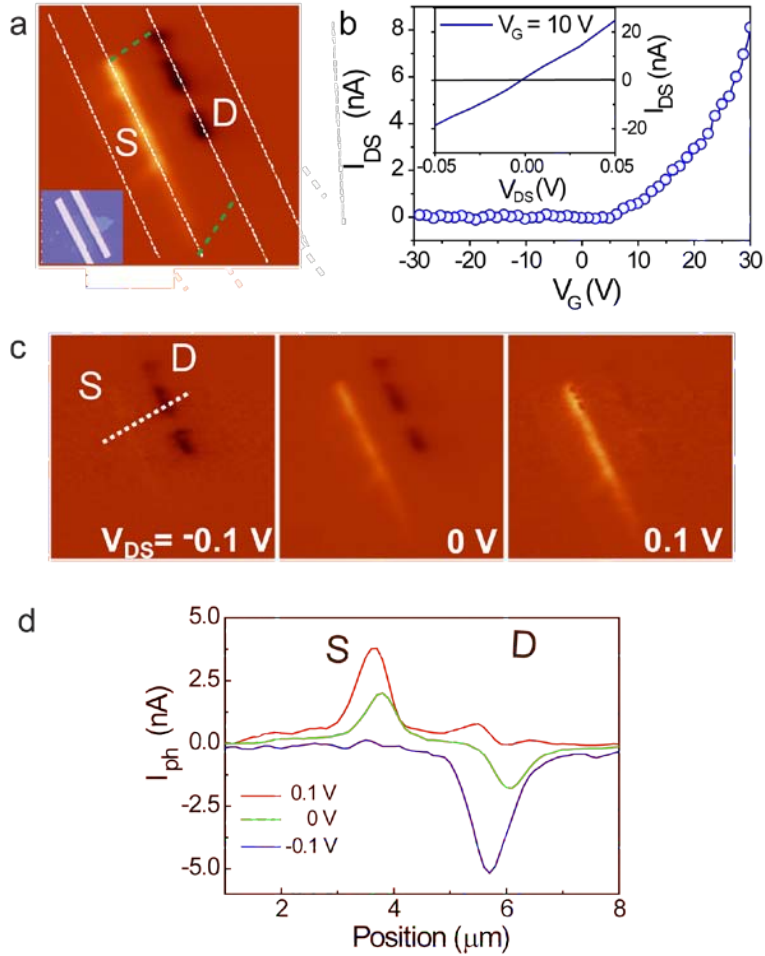

Figure S2: a) Spatial photocurrent image of a SL MoS$_2$ FET Inset: optical image of the SL device. Discontinuities in photocurrent are due to breaks in the flake. b) Electrical characteristics of the SL MoS$_2$ FET (L = 1.3 µm, W = 4 µm). Transfer curve is taken at $V_{DS}$ = 50 mV. The inset shows the linear output curve at low drain bias at $V_G$ = 10 V. c) Spatial photocurrent images of the SL FET at $V_{DS}$= 0.1, 0, and – 0.1 V. Positive (negative) photocurrent at S (D) is enhanced (suppressed) when applying positive bias, and vice versa. d) Photocurrent profiles from the images in c (white dash line) taken in regions without breaks in the film.

**Section S3: Carrier lifetime and diffusion lengths.**

To develop a more complete description of charge carrier transport characteristics in the FET, the minority carrier (hole) diffusion length $L_D$ was extracted from line profiles of the zero bias photocurrent. We convolved an exponential decay with the experimentally measured Gaussian beam width[2] (inset of Figure S3). The best fit (red line shown in Figure S3) yields $L_D \sim$ 0.22 μm. The results of this direct transport measurement agree with previously reported ultrafast optical measurements.[3] Furthermore, the field-effect mobility extracted from the transconductance, together with the diffusion length, can be used to estimate the minority carrier lifetime $\tau = \frac{L_D^2}{D}$, where $D = \mu k_B T/q$ is the diffusion coefficient, $q$ is electron charge, $k_B$ is the Boltzmann constant, and $T$ is temperature.[4] The calculated carrier lifetime $\tau \sim$ 2.4 ns also agrees with literature values.[5] The same analysis applied to the SL device yields a diffusion length of ~0.4 μm and a lifetime of ~60 ns.

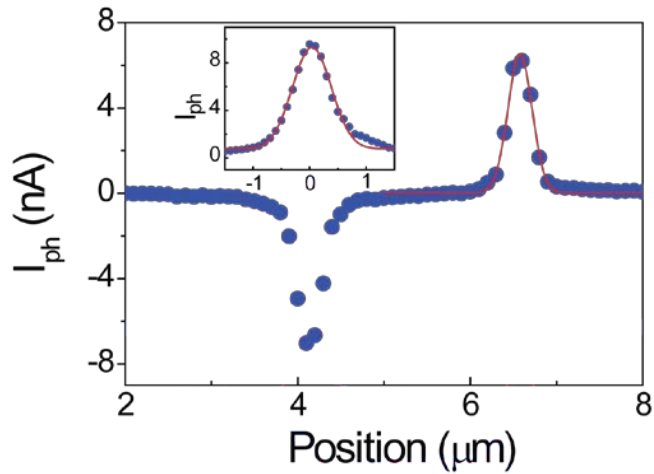

Figure S3: The photocurrent spatial profile (solid circles) is plotted along the channel. The solid red line represents the fitting using a convoluted Gaussian-exponential function. Inset shows the cross-section photocurrent profile (solid circles) taken perpendicular to a Si nanowire with diameter of 100 nm. The solid red line represents a Gaussian-peak fitting.

## Section S4: Gate bias dependence of photocurrents

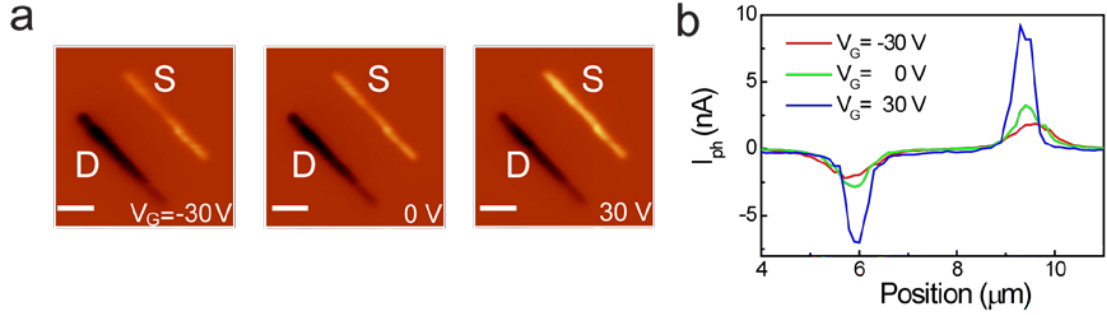

Figure S4. Gate bias dependence of photocurrents. a) Spatial photocurrent maps of a 4L MoS$_2$ FET at V$_{DS}$ = 0 V for V$_G$= -30, 0, and 30 V. b) Line profiles of photocurrents along the channel at V$_{DS}$ = 0 V for V$_G$= -30, 0, and 30 V.

## Section S5: Estimation of photoexcited and intrinsic carrier concentrations.

Under low injection conditions, the intrinsic carrier concentration is much greater than the photogenerated carrier concentration, and the photoconductivity is directly proportional to the illumination intensity. As the photogenerated carrier concentration becomes equal to or greater than the intrinsic carrier concentration, bimolecular recombination of photogenerated electrons and holes becomes significant, producing a sub-linear power dependence approaching $I_{ph} \propto P^{1/2}$.[6] Estimates of the densities of photoexcited and intrinsic carriers in MoS$_2$ were made using the parameters listed in Table I below. The excess carrier density ($\Delta n=\Delta p$) is given by the product of generation rate (G) and carrier lifetime (τ) ($\Delta n=\Delta p=G\cdot \tau$). The carrier lifetime is estimated as described in the main text, and the generation rate is obtained from absorbed power density ($P_{ab}$) via G=$P_{ab}$/hv. For a given excitation power density $P_{in}$, $P_{ab}=P_{in}(1-e^{-\alpha d})$, where α is the absorption coefficient and d is the thickness. Since $\alpha d \ll 1$ for thin MoS$_2$, $P_{ab} \sim P_{in}\alpha d$. For an incident power of 5 µW, the excess carrier density is ~$3 \times 10^{11}$ cm$^{-2}$. Next, the intrinsic carrier concentration of MoS$_2$ at V$_g$=0 V is estimated by the ratio of the conductance ($\sigma=qn\mu$) at two different V$_g$, where q is electron charge and µ is mobility (assumed to be constant within the range of V$_g$ considered). Assuming that the carrier concentration is roughly equal to the induced charge density ($C_{SiO_2} \times V_g$), where $C_{SiO_2}$ is the capacitance of SiO$_2$, the intrinsic carrier density at V$_g$= 0

V is estimated to be $\sim 10^{10}$ cm$^{-2}$. This is much less than the photoexcited carrier density ($3 \times 10^{11}$ cm$^{-2}$), which is consistent with our conclusion that bimolecular recombination accounts for the observed sub-linear photocurrent behavior. Moreover, the carrier concentration in MoS$_2$ approaches that of the photogenerated carrier density with increasing positive gate voltage, which reduces the nonlinearity of photocurrent with incident power (Figure 4b).

## Table I: key system parameters

| Parameter | Symbol | Value |
|---|---|---|
| Channel width | $W$ | 6.8 μm |
| Channel length | $L$ | 3.5 μm |
| SiO$_2$ thickness | $t$ | 300 (nm) |
| Numerical aperture | $NA$ | 0.9 |
| Excitation wavelength | $\lambda$ | 600 (nm) |
| Absorption coefficient | $\alpha$ | $2 \times 10^5$ (cm$^{-1}$) |
| MoS$_2$ thickness | $d$ | 2 (nm) |
| Carrier lifetime | $\tau$ | 2.4 (ns) |
| Generation rate | $G$ | $2 \times 10^5$ (cm$^{-1}$) |
| Capacitance of SiO$_2$ | $C_{SiO_2}$ | $7.2 \times 10^{10}$ (cm$^{-2}$/V) |

Note: Diameter (*D*) of focused beam is estimated by $D = 1.22 \cdot \lambda/NA$, and the capacitance of SiO$_2$ is calculated by $\varepsilon_0 \varepsilon_r / t$.

## Section S6: Time-resolved photoresponse

The photocurrent was converted into a voltage with a current preamplifier (DL Instruments 1211) and recorded with a digital sampling oscilloscope (Tektronix TDS 2014). The measured rise time (10% to 90%) of the photocurrent is ~30 µs for both 4L and SL device (Figure S5a,b) and a 10 kΩ test resistor biased with a 10 mV square wave (Figure S5c) using the same preamplifier settings. A comparison of the two signals shows that the device photocurrent response is limited by the current preamplifier bandwidth. The time-dependence of the photocurrent is consistent with a fast process such as interband excitation and charge carrier separation. In contrast, photocurrent transients associated with thermal effects in which electrons are in equilibrium with the lattice are expected to be several orders of magnitude slower (a few milliseconds to seconds).[7-8] We also note that the SPCM images were acquired with a lock-in at a modulation frequency of 1837 Hz, which would reject a lower frequency signal from contact junction heating if it were outside the bandwidth of 37.5 Hz.

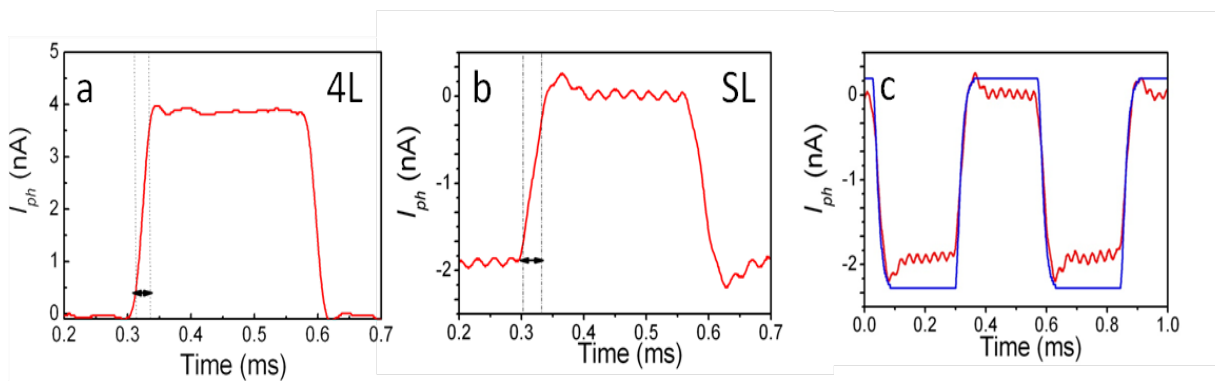

Figure S5: a) Time-resolved photocurrent of a 4L $MoS_2$ transistor. The rise time of the photocurrent is ~30 µs. b) Time-resolved photocurrent of a SL $MoS_2$ transistor. The rise time of the photocurrent is also ~30 µs. c) Time-resolved photocurrent (red) in comparison with the ac current through the test resistor (blue).